\documentclass[conference]{IEEEtran}
\IEEEoverridecommandlockouts

\usepackage{url}
\usepackage{soul}
\usepackage{amsmath,amsfonts}
\usepackage{algorithmic}
\usepackage[ruled,vlined ]{algorithm2e}
\usepackage{multirow}
\usepackage{array}
\usepackage{tabularx}
\usepackage{graphicx}
\usepackage{textcomp}
\usepackage{xcolor}
\usepackage{psfrag}
\usepackage{mathtools}
\usepackage{xspace}
\def\BibTeX{{\rm B\kern-.05em{\sc i\kern-.025em b}\kern-.08em
    T\kern-.1667em\lower.7ex\hbox{E}\kern-.125emX}}

\newcounter{challenge}

\newcounter{recommendation}

\newcounter{myposition}

\usepackage{balance}

\usepackage{listings}
\usepackage{xcolor}
 
\definecolor{codegreen}{rgb}{0,0.6,0}
\definecolor{codegray}{rgb}{0.5,0.5,0.5}
\definecolor{codepurple}{rgb}{0.58,0,0.82}
\definecolor{backcolour}{rgb}{0.95,0.95,0.92}
 
\lstdefinestyle{mystyle}{
    backgroundcolor=\color{backcolour},   
    commentstyle=\color{codegreen},
    keywordstyle=\color{magenta},
    numberstyle=\tiny\color{codegray},
    stringstyle=\color{codepurple},
    basicstyle=\ttfamily\footnotesize,
    breakatwhitespace=false,         
    breaklines=true,                 
    captionpos=b,                    
    keepspaces=true,                 
    numbers=left,                    
    numbersep=5pt,                  
    showspaces=false,                
    showstringspaces=false,
    showtabs=false,                  
    tabsize=2
}
 
\def\adityaIgnore#1{}

\begin{document}
\title{Attack Pattern Mining to Discover Hidden Threats to Industrial Control Systems   }

\author{
     \IEEEauthorblockN{Muhammad Azmi Umer$^1$ \thanks{1  Singapore University of Technology and Design, Singapore, \newline azmi\_umer@sutd.edu.sg}, Chuadhry Mujeeb Ahmed$^2$ \thanks{2 Newcastle University,  Newcastle, United Kingdom, \newline mujeeb.ahmed@newcastle.ac.uk}, Aditya Mathur$^3$ \thanks{3 Singapore University of Technology and Design, Singapore, \newline aditya\_mathur@sutd.edu.sg}, Muhammad Taha Jilani$^4$ \thanks{4 Bahria University, Pakistan, \newline mtahajillani.bukc@bahria.edu.pk }
     }
}

\maketitle

\begin{abstract}

This work focuses on validation of attack pattern mining in the context of Industrial Control System (ICS) security. A comprehensive security assessment of an ICS requires generating a large and variety of  attack patterns. For this purpose we have proposed a data driven technique to generate attack patterns for an ICS. The proposed technique has been used to generate over 100,000 attack patterns from data gathered from an operational  water treatment plant.  In this work we present a detailed case study to validate the attack patterns.

\end{abstract}


\def\comment#1{}

\section{Introduction}

 Industrial Control Systems (ICS) are at the heart of many critical infrastructure (CI), such as the electric power grid and water treatment plants. These systems control the physical processes  using  computing and communication elements, such as Programmable Logic Controllers (PLCs), Supervisory Control and Data Acquisition systems (SCADA), and communication networks~\cite{ScanningTheCycle_AsiaCCS2021}. While automation has simplified the monitoring and control of  CI, it has also exposed them to potential threats from malicious entities, as demonstrated by several  attacks~\cite{weinbergerStuxnet,ukraineBlackout,germanSteelMill}. Even air-gapping a system,  often considered a security measure, cannot guarantee protection against insider threats.

In this study, we refer to a technique\,\cite{umer2021attack} for automatically generating attacks based on data. The technique is used to generate attack patterns intended to move an ICS to an anomalous state.  An attack pattern, also referred to as an attack rule,  consists of states that are disturbed to move the plant to an anomalous state. A key advantage of using attack patterns to move a plant to an anomalous state is that they do not require design knowledge such as in\,\cite{sridhar_compsac_2016,Rocchetto2016}.  Attack patterns can generate a large set and variety of attacks practically impossible when generated manually\,\cite{sridhar_compsac_2016}. For example, only 36-attacks, generated manually, are reported in\,\cite{sridhar_compsac_2016} whereas 117,960 attacks were generated using attack patterns.   Since then we have conducted experiments to validate the attack patterns. The validation procedure used in the current work is significantly different from that used in\,\cite{umer2021attack} and is described in Section~\ref{sec:arm}. Results presented in this work attest to the  effectiveness of the attack patterns in moving the target plant to an anomalous state.

\noindent \emph{Motivation:} The idea of generating attack patterns arose from the challenges faced in the design and launch of attacks. In earlier work on testing anomaly detectors\,\cite{adepu_mathur_TDSC}, attacks were designed and launched manually on an operational water treatment plant, namely SWaT\,\cite{mathurTippenhauer}. This process was time consuming in that it required a thorough understanding of the multiple sub-processes and their interactions.   Even though SWaT is significantly smaller than a city-scale plant, the complexity of the treatment processes led to attacks designed that would not lead to the desired impact; and in some cases no impact at all. Thus, the manual process of designing and launching attacks became iterative and time consuming. The result was only a small number of attacks (36). The attack patterns method proposed and evaluated in this work overcomes the shortcomings mentioned above. Using the proposed method one can generate thousands of attacks that are launched automatically using an attack launcher.

\noindent \emph{Contributions:} The key contribution of this work lies in the validation of the approach for generating attack patterns and their intended use in critical infrastructure.  Attack patterns generated and validated here are available to researchers for validating anomaly detection methods.\footnote{\url{https://sites.google.com/view/crcsweb/attack-patterns}}

\noindent \emph{Organization:} The remainder of this paper is organized as follows. In Section~\ref{sec:relatedWork} we compare the proposed method with existing work on automated attack generation. A brief introduction to the SWaT testbed used for validation of attack patterns, and an illustrative example,  are in Section~\ref{sec:SWaT}. A method for generating attack patterns is summarized in Section~\ref{sec:arm}. The case study for validating attack patterns generated using our method, and results obtained, are  described in Section~\ref{sec:validation}. Outcomes of the case study is further analyzed in Section~\ref{sec:discussion}. Conclusions from this study are in Section~\ref{sec:conclusions}.

\iffalse
\section{Introduction}
% Introduction to the CPS and the attack detection problem

\fi

\section{Related Work}
\label{sec:relatedWork}

Previous efforts to manually create attacks faced scalability issues and were constrained by the expertise of the domain expert. However, the application of machine learning for automatic attack generation has gained significant interest~\cite{lin2019idsgan_IDSGAN}. Much of the research has concentrated on adversarial learning aimed at producing evasive attacks. For instance, in some studies~\cite{zizzo2020adversarial_trustcom2020_Imperial}, researchers successfully generated stealthy attacks tailored for specific detectors using the SWaT attack dataset~\cite{sridhar_dataset_paper}. Another study~\cite{kravchik2020poisoning_BGU_AsafShabtai} utilized the SWaT dataset to incrementally perturb input samples in the training process to evade detection during testing. They selected seven attacks from the dataset and developed stealthy attacks for a threshold-based detector operating on residual signals. These approaches leverage publicly available attack datasets~\cite{sridhar_dataset_paper} and adapt them to create a limited number of stealthy attacks for specific detection methods. In~\cite{feng2017deep_3,lin2019idsgan_IDSGAN}, Generative Adversarial Networks (GANs) were used to train anomaly detector classifiers and generate undetectable malicious sensor measurements. The study reported in \cite{JIA2021100452} explored adversarial attacks against Industrial Control Systems (ICS). They used gradient-based methods to evade the RNN-based anomaly detectors in WADI~\cite{wadi2017} and SWaT. The authors in~\cite{erba_nils_2} devised attacks to evade autoencoders in water distribution systems by creating two distinct spoofed sensor values, one for PLC and another one for the detector. While these studies primarily target specific classifiers and try to manipulate either sensor or actuator data, our proposed approach does not have these limitations. We have used a limited attack dataset to autonomously generate attack patterns.

In addition to adversarial learning, methods exist in the literature to generate synthetic attack data. A study reported in~\cite{pham2014generating_attacks} used normal data to create anomalous data by sampling out-of-distribution examples. While this method can produce anomalous data, it does not necessarily generate attack data, making it less useful for security testing. Several methods reported can produce data samples from the minority class, i.e., the anomalous class~\cite{steinbusscalibration_SMOTE_others}, such as Synthetic Minority Over-sampling Technique (SMOTE)~\cite{chawla2002smote}. It generates synthetic data points from the minority class. While SMOTE effectively balances datasets for machine learning, it cannot enrich the dataset with novel scenarios rather it creates synthetic variations using existing anomalous samples. Similarly,  work on generating anomalous data on power grids~\cite{turowski2022modeling_anomalousDataPowerGrid} also proposed a technique to generate synthetic anomalous data similar to observed real anomalies. However, it again generates more samples of what is already known rather than generating new attack patterns. In contrast, our proposed technique generates many attack patterns never been seen before and are challenging to derive manually. The objective of our technique is to understand the consequences, i.e., plant damage, of applying attack patterns designed by process experts and to learn the different ways such damage can be inflicted.

\section{The SWaT testbed and Attack Patterns}
\label{sec:SWaT}

\subsection{SWAT Testbed}
\label{sec:swat}
% an explanation on the task of attack detection in  a water system is given here

The Secure Water Treatment (SWaT) system, a full-scale industrial model of a water treatment facility\,\cite{mathurTippenhauer}. This system is often used for validating ICS defense strategies\,\cite{swatDataset}. %A visual representation of the SWaT facility can be seen in Figure~\ref{fig:SWaT_Testbed_Pic}. 
The system is capable of generating 5-gallons/minute of treated water.  SWaT is a multi-stage distributed control system, with six distinct stages. Each stage is equipped with a variety of sensors and actuators. 
In the first stage, raw water is received from an external source. The second stage involves chemical dosing  based on readings from the water quality sensors. The third stage carries out ultrafiltration. The fourth stage involves dechlorination before  water is sent to the reverse osmosis units in the fifth stage. The sixth and final stage distributes the treated water and also carries out backwash to clean the ultrafiltration unit. Communication between sensors/actuators and PLCs is facilitated by a Level\,0 network, while inter-PLC communication occurs over a Level\,1 network.

\begin{figure}
    \centering
    \includegraphics[width=0.8\linewidth,height=3.25cm]{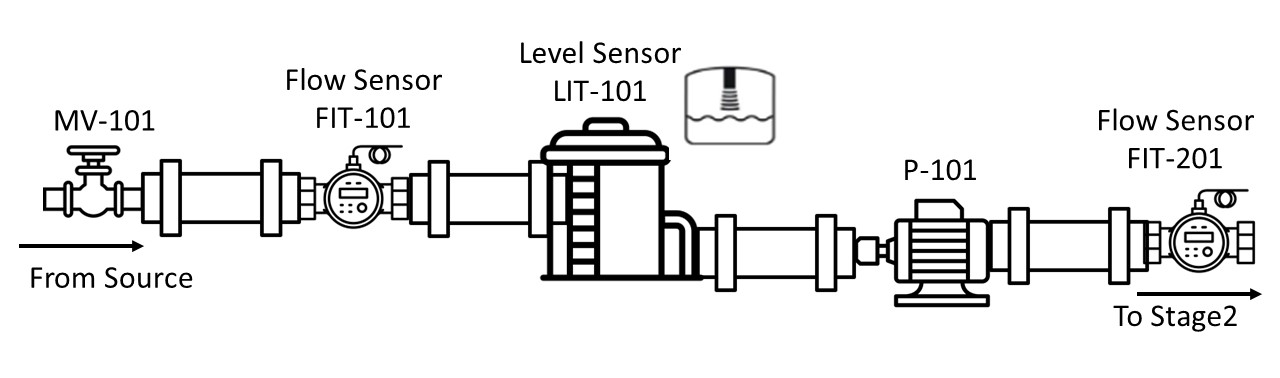}
    \caption{Stage~1 of the SWaT Testbed.}
    \label{fig:Stage1_SWaT}
\end{figure}
 
\subsection{Attack patterns: an illustrative example} 

\begin{figure}
\centering
\includegraphics[width=0.7\linewidth,height=6cm]{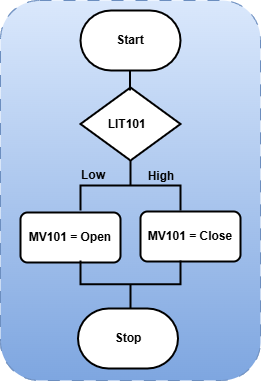}
\caption{An example of control logic implemented in the first stage of SWaT}
\label{fig:controlLogicStage1}
\end{figure}

Figure~\ref{fig:Stage1_SWaT} depicts the first stage of SWaT. There is a raw water storage tank with a level sensor (LIT-101)  positioned at the top. An inlet motorized valve (MV-101) and flow meter (FIT-101) are installed on the inlet pipe of the raw water tank. The tank's outlet is connected to a pump (P-101) that is further connected to a flow sensor (FIT-201). A motorized valve (MV-201) controls the outflow towards the second stage of SWaT. The tank's primary function is to store raw water for subsequent purification. With these critical components identified, let us examine how this process is controlled by a PLC. Figure \ref{fig:controlLogicStage1} provides an example of the control logic implemented in the PLC.

Consider the following attack pattern intended to move  SWaT Stage\,1 to an anomalous state.
\vskip0.1in
\begin{tabular}{lll}
FIT101$>$0.5,&MV201=Close,& MV302=Open, \\  MV303=Close,& P302=Off\\
$\rightarrow$ MV101=Open
\end{tabular} 
The pattern above consists of two parts: a condition followed by an action. Thus, the pattern can be read as follows: ``Open valve MV101 when flow rate, indicated by FIT101, is above 0.5, valves MV201 and MV303 are closed, valve MV302 is open, and pump P302 is stopped.'' The above attack pattern was used to launch an attack on the SWaT testbed. It led to overflow of tank T101. Thus, the attack pattern was effective in realizing its intention of moving the plant to an anomalous state. Examples of similar  patterns, generated for SWaT,  are listed in Section~\ref{sec:validation}.

\comment{The tank has predefined operational limits for holding water: the Low\_Level represents the minimum threshold below which stage~1 cannot fulfill the water demands of further stages,
while the High\_Level is the maximum allowable water level in the tank before an overflow occurs. As shown in figure \ref{fig:controlLogicStage1}, the control system is designed in such a way that it shuts off the inlet supply (MV-101) when the level is high. This strategy prevents tank overflow and subsequent flooding. A malicious actor can cause tank overflow through various strategies. One approach is to manipulate MV-101 such that it remains open even when the high level is reached. This attack, which is solely targeting the MV-101, would be relatively easy to detect as the tank level readings on the SCADA system would exceed the expected values. A more sophisticated attack could involve manipulating both the level sensor (LIT-101) and motorized valve (MV-101). By keeping MV-101 open and falsifying level sensor data to simulate normal levels, an attacker can overflow the tank. This method would evade the alarms that are triggered solely based on level sensor data, therefore making the SCADA system incapable of detecting these intrusions. The complexity of the attack increases by compromising multiple components of the system. While Stage 1 appears relatively simple, however, it encompasses seven different sensor/ actuators, resulting in a comprehensive array of potential attack configurations. Given the considerably greater complexity of subsequent stages of SWaT and real-world systems, which comprise hundreds of devices, an exhaustive manual identification of all feasible attacks becomes impractical. Therefore, the proposed study has introduced a data-driven and automated approach for generating potential attack patterns for the ICS.}

\begin{figure}
\centering
\includegraphics[width=1.0\linewidth,height=12.25cm]{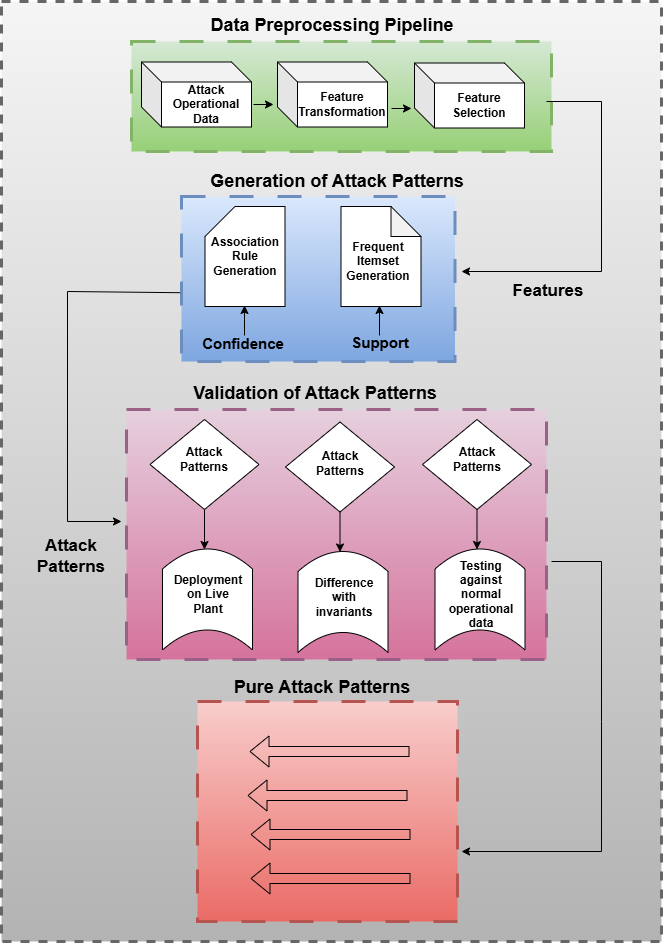}
\caption{Generation of Attack Patterns using ARM.}
\label{fig:genval}
\end{figure}

\section{Generating Attacks via Association Rule Mining}
\label{sec:arm}
An important challenge in the anomaly detection problem is the limited availability of diverse attack data. The method proposed in the current study addresses this issue by laying foundation upon the attack data designed by the domain experts ~\cite{sridhar_dataset_paper}. We used this dataset to create a comprehensive set of potential attack scenarios. The attack data was collected over four days of continuous operation at the SWaT testbed. SUTD engineers and researchers considered a comprehensive set of attack scenarios to simulate the attacker behaviors, and their intended goals. For example, the tank overflow attack, which has been described earlier, is an example of this approach. Essentially, an attack model focuses on the identification of a specific sensor or actuator to target and subsequently its implication on the overall system's physical processes. This complete exercise yielded a total of 36 distinct attack patterns. Obviously it is not comprehensive, however this dataset has potential significance for developing machine learning-based security solutions. We used the aforementioned work as the foundation for our approach. Using this attack dataset, our approach generated a large number of attack patterns. Since each of the original 36 attacks was designed with a specific objective, e.g. causing a tank overflow. Therefore, by analyzing these attack dataset, we can understand the underlying attacker intent and can explore all potential methods to achieve similar goals. To uncover these potential methods, Association Rule Mining (ARM) technique \cite{agrawal1993mining}, is applied to generate potential attacks. ARM is an unsupervised machine learning method that uses data itself to generate the association rules present in the data. Though, ARM is commonly used in market basket analysis but it has demonstrated significant effectiveness in intrusion detection \cite{ahmed2021machine}. The proposed study has used ARM to generate attacks on the secure water treatment plant (SWaT). The overall procedure is described in Algorithm \ref{alg:in}. The SWaT's attack data \cite{swatDataset} was used to generate the attack patterns. The following type of attack was generated using the proposed method:

\begin{equation}
\label{eq:in}
    X\;\Longrightarrow\; Y
\end{equation}

There are various algorithms for the implementation of Association Rule Mining (ARM). However, in the proposed study, we have utilized the FP-growth algorithm. The Orange-Associate library contains the implementation of aforementioned algorithm in Python. ARM involves a two-step process: First is the identification of frequent itemsets and the second is the derivation of association rules from these frequent itemsets. The overall process is visualized in the Figure \ref{fig:genval}.

\subsection{Frequent Itemsets}

An itemset can consist of a value of single attribute or it can be a combination of values of multiple attributes within a dataset. If an itemset meets the predefined minimum support threshold, then it is classified as a frequent itemset.

\subsubsection{Support:} 
The support of an itemset 'I' within a dataset 'D' is determined by counting the number of transactions or rows 'r' in the dataset 'D' that include the itemset 'I'.

\begin{equation}
\textstyle{S}(I) = \frac{|r \in D; I \in r|} {|D|}
\end{equation}

\subsection{Association Rules}
Association rules are derived from frequent itemsets, as described in Equation \ref{eq:in}. A rule must meet a specified minimum confidence threshold for being considered as a valid association rule.

\subsubsection{Confidence:}
The rule described in Equation \ref{eq:in} consists of two components i.e. the antecedent (X), which is on the left side of the implication symbol ($\Longrightarrow$), and the consequent (Y), which is on the right side of the implication symbol. The confidence of any rule is calculated by dividing the support of both antecedent and consequent combined by the support of the antecedent alone.

\begin{equation}
\textstyle{C}(X \;\Longrightarrow\; Y) = \frac{S(X \cup Y)} {S(X)}
\end{equation}

\begin{algorithm}
\caption{Generation of Attack Patterns using ARM}
\label{alg:in}
\begin{algorithmic}[1]
\STATE Capture network packets.

\STATE Decode packets to extract sensor and actuator data.

\STATE Store the state information from sensor and actuator data.

\STATE Transfer the extracted state information to the historian.

\STATE Transform the features of the dataset, and select the most relevant features.

\STATE Generate frequent itemsets (i.e. frequent patterns in the dataset).

\STATE Generate potential attack patterns (i.e. association rules) from frequent itemsets generated in the earlier step.

\STATE Validate the generated attacks by comparing them against the normal system behavior

\end{algorithmic}
\end{algorithm}

\subsection{Feature Engineering for the Generation of Attack Patterns}
\begin{figure}
\centering
\includegraphics[width=1.0\linewidth,height=6cm]{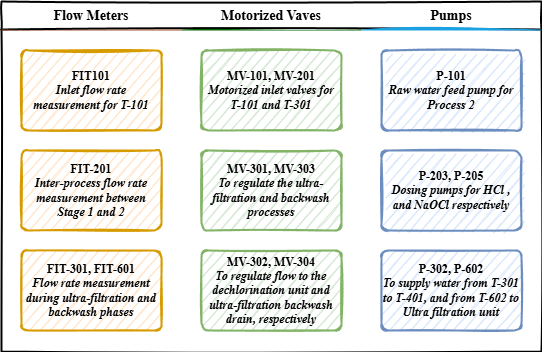}
\caption{Attributes selected for ARM}
\label{fig:attrDescrp}
\end{figure}

The SWaT dataset contains 51 attributes that represent various sensors and actuators. These attributes can take on binary i.e. On or Off in the case of pumps, ternary i.e., Open, Closed, or Transition in the case of a motorized valve, and continuous values i.e. level sensors in the case of tanks. The Association Rule Mining (ARM) can only process binary data, therefore all attributes in the dataset must be transformed into a binary valued attribute for compatibility with the ARM algorithm. The transformation of attributes into the binary valued attribute resulted in creating some useless data for ARM. For example, some attributes were reduced to a single binary value, rendering them useless for analysis. Therefore, these attributes were removed from the dataset. To ensure a proper comparison with the baseline research \cite{UMER_Azmi_2020}, we selected the same 15 attributes for generating attack patterns as done in \cite{UMER_Azmi_2020}. These attributes are described in Figure \ref{fig:attrDescrp}. This allowed the direct validation of our generated attack patterns against established normal behavior profiles in \cite{UMER_Azmi_2020}.

\subsubsection{Feature Transformation}
\begin{figure}
\centering
\includegraphics[width=0.7\linewidth,height=7cm]{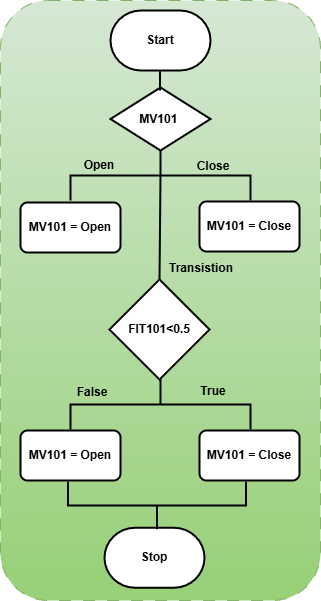}
\caption{Transformation of ternary valued attribute into binary valued attribute}
\label{fig:AttributecontrolLogic}
\end{figure}

The dataset contains attributes such as motorized valves that have three states: Close (1), Open (2), and Transition (0). This makes them the ternary valued attributes. The duration of transition state is very small, i.e., less than 10 secs. To transform these attributes into binary ones, the transition state was changed to either the Open or the Close state. This was done by analyzing the respective FIT. If the FIT is below 0.5 for the respective transition state of motorized valve, then it was classified as 'Closed'; otherwise, it was marked as 'Open'. This process is shown in Figure \ref{fig:AttributecontrolLogic}. The FITs which are the real-valued attribute were converted into binary valued attribute. A FIT value of 0.5 or higher indicates the presence of flow, while the value lower than 0.5 implies that there is no flow.

\subsubsection{Significance of Support Threshold in the Generation of Association Rules}
The correct choice of support threshold has significance in rule generation. A high support threshold limits the number of generated rules, therefore it creates the possibility to overlook some significant patterns in the data. Similarly, a low support threshold can lead to a very large number of rules. For example, in the SWaT dataset, there is an attribute, Pump P602, which runs 
 only for 3164 seconds i.e. only 3164 out of 410400 total transactions. To identify patterns associated with P602 when running, the support threshold must be lowered to a minimal level of 0.77\% i.e. 3164/410400.

%\section{Evaluation and Discussion}
\newcolumntype{C}[1]{>{\centering\let\newline\\\arraybackslash\hspace{0pt}}m{#1}}

\begin{table*}[]
\caption{Validation on Operational Testbed (1 of 2)}
\label{tab:validation1}
\begin{tabular}{|p{0.3in} C{2.2in} p{0.8in} p{0.5in} p{0.5in} p{2in}|}
\hline 
 {\textbf{ S.No.}} & \textbf{Attack Rule}                                                                                             & \textbf{State Prefix}       & \textbf{Start Time} & \textbf{End Time} & \textbf{Impact}                                                                                                                                                            \\ \hline
1                                  & MV101=Close, MV303=Close, MV304=Open, P302=Off --\textgreater P205=Off                                                 & N/A                & {11:03}      & 11:10    & RO is not receiving the water                                                                                                                                                                                                          \\ 
\hline
2                                  & P203=Off, MV101=Close, P602=Off, FIT201$<$0.5 --\textgreater FIT601$<$0.5                                               & N/A                & N/A  &  N/A    &  Plant was already operating as per the attack rule                                                                                                                                                                 \\ 
\hline
3                                  & FIT101$>$0.5, MV304=Close, P101=Off, P302=On --\textgreater 
 MV303=Close                                                 & MV302=Open or MV304=Open &  11:25      & 11:30    & Water in RO is increasing continuously and can overflow the RO   Feed Tank                                                                                        \\ 
\hline
4                                  & FIT101$>$0.5, MV201=Close, MV302=Open, MV303=Close, P302=Off --\textgreater MV101=Open                                      & N/A                & 11:34      & 11:41    & Level of water in the raw water tank is continuously increasing   and could overflow the tank                                                                        \\ 
\hline
5                                  & P302=Off, MV101=Close, MV304=Close, FIT201$<$0.5 --\textgreater P602=Off                                                & N/A                & N/A   & N/A     &       Plant was already operating as per the attack rule                                                                                                                                                             \\ 
\hline
6                                  & P205=On, P302=Off, MV301=Close --\textgreater  P203=On                                                           & MV201=1, P101=On    & 11:45      & 11:49    & Causing overdosing of chemicals                                                                                                                                   \\ 
\hline
7                                  & P101=Off, FIT201$<$0.5, P205=Off, FIT301$<$0.5, MV302=Open --\textgreater MV101=Open                                      & N/A                & 09:51     & 09:55    & No impact observed                                                                       \\ 
\hline
8                                  & MV302=Close, MV101=Open, MV303=Close, P205=On --\textgreater P602=Off                                                 & MV201=Open, P101=On    & 09:59      & 10:01    & Reading in ORP meter is increasing and Level of water in the raw   water tank is continuously increasing and could overflow the tank                                 \\ 
\hline
9                                & MV101=Open, MV201=Close, MV302=Open, MV303=Close, P302=Off --\textgreater FIT101$>$0.5                                      & N/A                & 10:03      & 10:05    & Level of water in the raw water tank is continuously increasing   and could overflow the tank                                                                        \\
\hline
10                                 & P203=Off, MV101=Close, MV303=Close, FIT301$<$0.5 --\textgreater FIT201$<$0.5                                              & N/A                & N/A & N/A       &       Plant was already operating as per the attack rule                                                                                                                                                             \\
\hline
11                                 & P101=Off, FIT201$<$0.5, MV201=Close, P203=Off, P205=Off, FIT301$<$0.5, MV301=Close,   MV302=Open, FIT601$<$0.5 --\textgreater P302=Off & N/A                & 10:09      & 10:12    & Backwash unable to start                                                                                                                                          \\ 
\hline
12                                 & FIT101$>$0.5, P101=Off, MV201=Open, MV301=Close, MV304=Close, P302=On --\textgreater   FIT301$>$0.5                           & N/A                & 10:14      & 10:17    & The UF membrane could be overused, Level of RO Feed tank is   increasing faster                                                                                  \\ 
\hline
13                                 & MV201=Close, MV302=Close, MV101=Close, MV301=Close --\textgreater P203=Off                                                & N/A                & 10:20      & 10:22    & Its preventing the filling of RO Feed water tank                                                                                                                  \\
\hline
14                                 & MV302=Open, P205=Off, FIT301$<$0.5 --\textgreater P203=Off                                                         & N/A                & 10:24      & 10:26    & Its preventing the filling of RO Feed water tank                                                                                                                  \\ 
\hline
15                                 & MV302=Close, P602=Off, P205=On, MV301=Close --\textgreater P203=On                                                  & MV201=Open, P101=On    & 10:29      & 10:32    & Value of pH meter is decreasing and value of ORP meter is   increasing                                                                                                       \\ \hline

\end{tabular}

\end{table*}

\begin{table*}[]
\caption{Validation on Operational Testbed (2 of 2)}
\label{tab:validation2}
\begin{tabular}{|p{0.3in} C{2.2in} p{0.8in} p{0.5in} p{0.5in} p{2in}|}
\hline 
 {\textbf{ S.No.}} & \textbf{Attack Rule}                                                                                             & \textbf{State Prefix}       & \textbf{Start Time} & \textbf{End Time} & \textbf{Impact}                                                                                                                                                            \\ \hline
16                                 & MV201=Close, MV302=Open, FIT301$<$0.5, FIT601$<$0.5 --\textgreater P203=Off                                              & N/A                & 10:34      & 10:37    & Its preventing the filling of RO Feed water tank                                                                                                                  \\ \hline
17                                 & MV201=Open, MV304=Close, P101=Off, FIT201$<$0.5 --\textgreater MV301=Close                                               & N/A                & 10:41      & 10:44    & P102 started and soon stopped as the water in UF feed water   tank was already high                                                                                \\ \hline 
18                                 & MV303=Open, MV101=Open, FIT301$<$0.5, FIT201$>$0.5 --\textgreater MV302=Close                                             & N/A                & 10:47      & 10:50    & Level of water in the raw water tank is continuously increasing   and could overflow the tank, and It's preventing the filling of RO Feed water   tank                \\ \hline
19                                 & MV302=Close, FIT301$<$0.5, MV301=Open, FIT201$>$0.5 --\textgreater MV201=Open                                             & N/A                & 10:53     & 10:56    & No impact observed                                                                                                                                                \\ \hline
20                                 & MV201=Open, P205=On, MV301=Close, P302=Off, FIT601$<$0.5, P602=Off --\textgreater   P203=On                              & P101=On             & 11:05      & 11:08    & UF Feed water tank is increasing above high, pH is decreasing,   and ORP is increasing                                                                            \\
\hline
21                                 & FIT201$>$0.5, P205=On, MV302=Close, P302=Off, FIT601$<$0.5 --\textgreater P203=On                                       & MV201=Open, P101=On    & 11:10      & 11:13    & Water from UF Feed tank is going to drain instead of RO Feed   Water Tank, UF Feed water tank is increasing above high, pH is decreasing,   and ORP is increasing \\
\hline
22                                 & FIT101$>$0.5, MV101=Open, MV301=Open, FIT201$>$0.5 --\textgreater MV201=Open                                             & N/A                & 11:16      & 11:19    & P203 turned on even though there is no water flow in stage 2,   level of water tank is increasing                                                                 \\ \hline
23                                 & MV201=Open, FIT601$<$0.5, MV303=Close, P205=On --\textgreater P602=Off                                                & P101=On             & 11:22      & 11:25    & ORP is increasing and UF Feed Level is increasing fastly                                                                                                          \\ \hline
24                                 & FIT201$>$0.5, P205=On, FIT301$<$0.5, MV302=Close, MV304=Close, FIT601$<$0.5 --\textgreater   P203=On                          & MV201=Open, P101=On    & 11:28      & 11:31    & ORP is increasing and pH is decreasing, level of water in RO   feed tank is increasing                                                                            \\ \hline
\end{tabular}

\end{table*}

\section{Validation of Attack Patterns}
\label{sec:validation}
It is important that the found attack rules are further tested for success in moving the target plant to an anomalous state. Since the attack rules are obtained from the attack data which also had some normal instances, therefore a test is designed to check the generated attack patterns. We have used the normal data from the SWaT testbed and then examined all patterns in this data for a combination of inputs and outputs that match any of those attack rules. The goal is to discover how many of these attack patterns can be found in the normal data and hence labeled as the ``wrong attack." Therefore,  a three-way strategy is adopted to evaluate the attack patterns. First, we evaluate an attack pattern against the invariants for the same testbed. The second is to validate it against the dataset of the plant. The third is to validate the attack pattern by launching it on the operational testbed. 

\subsection{Validation using  invariants}
We reproduced the experiments conducted in\,\cite{UMER_Azmi_2020} using a data-centric approach to validate the generated attack patterns. A primary issue was the presence of normal data samples in the attack dataset, which hindered the accurate identification of attacks within the generated attack patterns. To overcome this issue, we employed seven days of normal operational data from SWaT to generate invariants which were representing the normal behaviour of the plant. These invariants were generated using the ARM. The feature selection was performed on binary transformed 
attributes to derive the same features as in\,\cite{UMER_Azmi_2020}. Considering the full size of data rows as 410,400 we used the support of 1/410400 to generate the frequent itemsets from the transformed data. A 100\% confidence level was set to mine association rules from the frequent itemsets. The attacks generated by attack rules were compared with the invariant rules generated using ARM on normal data from SWaT. 
Specifically, we created a mathematical set of all the attack patterns, which is denoted by A, where each entry is an attack pattern. Similarly, we created another mathematical set based on invariant rules, denoted by B, where each entry represents an invariants of the SWaT. We took the difference of these two sets to obtain another set, denoted by C i.e. set C ($C = A - B$). Set C contains the attack patterns which are not the invariants of SWaT.
The number of attack patterns before validation in set A was 835,020. The number of invariants in set B was 4,502,209. The set C which represents the attack patterns after validation against invariants consists of 117,960 attack patterns. The number of invariants that were invalidated is 717,060. It means that 717,060 out of 835,020 i.e. 85.87\% attack patterns were invalidated by the proposed validation technique. Here, the question arises as, to why such a large number of invariants were invalidated. Is this the problem with the proposed approach? The answer is that the proposed approach learns or extracts the patterns from the data. The patterns could consist of two or more attributes. However, in the attacked dataset, the attacks were launched on some specific attributes. Only the attacked attributes and the attributes that have a relationship with them get disturbed due to the attack not all the attributes. Therefore, there are a lot of attributes in the attack data that represent the normal behavior of the plant. The proposed approach also extracts the patterns from these normal attributes. This is the reason for a large number of invalidated attack patterns because they actually represented the true behaviour of the plant.  
%\textcolor{red}{In this subsection we have identified the steps taken to do this...what are results, are those discussed separately? like the next subsection is discussing the results.... this one was abruptly ended}

\begin{table*}[]
    \centering
    \caption{Validation on different datasets of SWaT}
    \label{tab:datasetValidation}
    %\resizebox{8.4cm}{0.9cm}{
    \begin{tabular}{|c|c|c|c|}
    \hline
        \textbf{Dataset Year} & \textbf{Dataset Size} & \textbf{No. of Attack Rules in Normal Data} & \textbf{False Attack (\%)}  \\
        \hline
         2015 & 410400 rows & 3026 & 2.56\%
         \\ \hline
         
         2019 & 14996 rows & 2472 & 2.09\% 
         \\ \hline
         
         2020 & 18000 rows & 1707 & 1.44\%
         \\ \hline
    \end{tabular}
    %}
\end{table*}

\subsection{Validation using the dataset of the operational testbed}
We generated 117,960 attack patterns using our proposed methodology. All these attack patterns were validated on the complete dataset of the SWaT testbed which was collected during the normal operation of the plant in the year 2015. This dataset contains the record of plant activities at the interval of each second. We checked that if the attack rule generated by us was found in any interval then we considered it as an invalidated or wrong attack rule. Out of 117,960 attack rules, only 3,026 attack rules, i.e., 2.56\% were found in the normal operation of the plant as mentioned in Table \ref{tab:datasetValidation}. This means that these 3,026 attack rules should not be considered attack rules because this behaviour was present in the normal operation of the plant. We also validated the attack rules on the other datasets of SWaT which were collected in the year 2019 and 2020. In the year 2019 dataset, 2,472 out of 117,960 attack rules, i.e., 2.09\% were found in the normal operation of the plant. Similarly, In the 2020 dataset, 1,707 out of 117,960 attack rules, i.e., 1.44\% were found in the normal operation of the plant. The difference in wrong or invalidated attack rules in different datasets could be due to the sizes of these datasets.

\subsection{Live Validation using the Operational Testbed}
We also validated a random subset of attack rules by launching them live on the operational testbed. The details are given in the Table \ref{tab:validation1} and \ref{tab:validation2}. We randomly chose and deployed 24 different attack rules on the operational testbed at specific time intervals. Out of these 24 attack rules, fourteen rules with antecedent size 4, four with antecedent size 5, three with antecedent size 6, two with antecedent size 3, and one with antecedent size 9 were selected to ensure a variety in the selected rules. The time interval was from two minutes to a maximum of seven minutes. We intentionally avoided a longer time window due to the time limitation and to avoid any major fault in the testbed. In some cases, we have to set some state prefixes to deploy the attack rule otherwise, it would not be possible to launch the attack rule due to the interlock properties of various sensors and actuators. For example, the attack rule mentioned in S.No. 23 of Table \ref{tab:validation2} was not possible to launch unless we have started the Pump P101. We assess the impact of each attack on the testbed. The attack rule mentioned earlier lasted for three minutes which disturbed the Oxidation Reduction Potential (ORP) value and caused it to increase. Moreover, it also caused a rapid increase in the feed level of the Ultra Filtration (UF) tank. The attack rule mentioned in S.No. 9 of Table \ref{tab:validation1} caused a continuous increase in the water level of raw water. The attack lasted for only two minutes because it may overflow the tank. The attack rule no. 22 caused the pump P203 to turn on despite the fact that there was no water flow in stage 2 of the SWaT. This attack essentially added unnecessary chemicals into the water pipeline potentially making it hazardous for human consumption.  Apart from this, there were a few attack rules which were already representing the normal behaviour of the plant. It means that they are not the correct attacks rather it was the normal behavior of the plant as mentioned in attack rules no. 2, 5, and 10. There were also some attack rules given in S.No.7 of Table \ref{tab:validation1} and S.No. 19 of Table \ref{tab:validation2} that does not cause any impact on the plant. These attack rules were launched for three to four minutes but we were not able to assess any impact in this time window. The details of all attack rules are available in Table \ref{tab:validation1} and Table \ref{tab:validation2}. We can see that 19 out of 24 attacks resulted on real impact confirming them being attacks. Two attacks did not show any impact on the plant during the duration of these attacks while three attack rules resulted in false attacks.

\begin{figure}[tbh]
    \centering
    \includegraphics[width=0.9\linewidth,height=5cm]{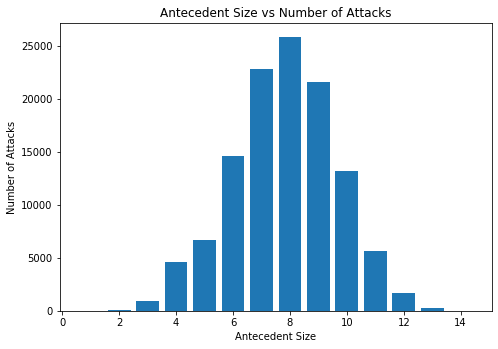}
    \caption{Impact of Antecedent size on the number of Attacks.}
    \label{fig:antecedent}
\end{figure}

\begin{figure*}[tbh]
    \centering
    \includegraphics[width=0.6\linewidth,height=7cm]{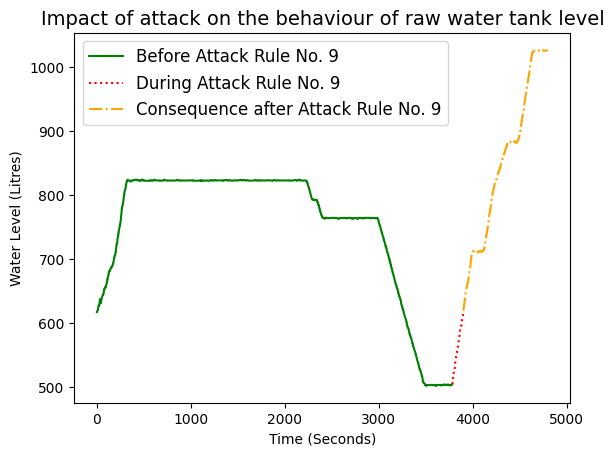}
    \caption{Impact of Attack on the Behaviour of Raw Water Level.}
    \label{fig:attackImpact}
\end{figure*}

\begin{figure*}[tbh]
    \centering
    \includegraphics[width=\linewidth,height=5.5cm]{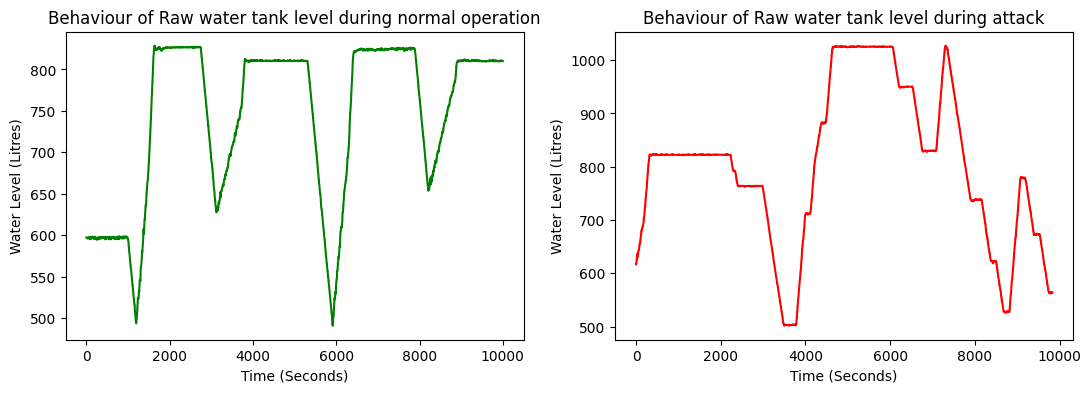}
    \caption{Behaviour of raw water Level during normal operation and under attacked scenario.}
    \label{fig:rawWater}
\end{figure*}

\begin{figure*}[h]
    \centering
    \includegraphics[width=\linewidth,height=5.5cm]{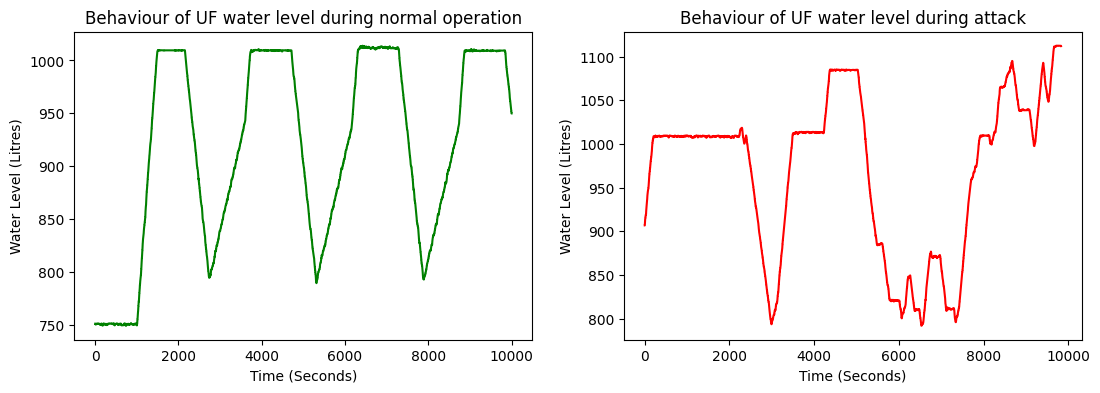}
    \caption{Behaviour of UF water Level during normal operation and under attacked scenario.}
    \label{fig:ufWater}
\end{figure*}

\section{Analysis}
\label{sec:discussion}
The 36 attacks generated by the domain experts in~\cite{sridhar_dataset_paper} served as the foundation of our approach to generate attack patterns.  Local and global attack patterns were generated using this approach. Local attack patterns target the individual processes in an ICS, whereas global attack patterns target multiple processes. These attack patterns involved the combination of 2 to 14 different sensors/actuators from a total of 15\,sensors/actuators considered.

As shown in Figure \ref{fig:antecedent}, the distribution of attack patterns follows a normal curve based on the number of generated attack patterns and the number of sensors/actuators involved. This analysis reveals that there is a small set of attack patterns when using only one or two devices. However, this set increases rapidly as more devices are incorporated. Though including additional sensors and actuators can increase the potential harm to an ICS,  there might be no underlying physical relationship among all the devices considered. Based on our findings, and for this case study, the optimal number of devices for maximizing attack scenarios is eight. An attack is launched by changing sensor measurements and states of the actuators as per the antecedent and consequent of the attack rule. These attack rules, or attack patterns, have significance for signature-based intrusion detection systems in that they provide attack data for industrial processes. The proposed method has generated over 110,000 attack patterns for SWaT. We intend to make our findings and dataset publicly available for researchers to use for further analysis on SWaT.

As mentioned earlier, the three-way validation strategy used invariants of  SWaT, multiple datasets, and live validation on the SWaT testbed. During live validation, the impact of an attack rule was studied on the individual sensors when the attack was active and upon its removal. For example,  Figure \ref{fig:attackImpact} shows the impact of an attack on the water level in the raw water tank T101. The green color represents the water level prior to attack launch  and red represents the water level when the attack was in progress. A sharp increase is observed in the level of water during the attack. The orange colour represents the water level when the attack is removed.   Though the attack has been removed, its impact persists in the form of continuous rise in water level which can potentially lead to tank overflow. The maximum allowed water level in the tank is 800mm though in  Figure\,\ref{fig:attackImpact} tthe level has crossed the 1000mm mark. Thus, we also studied the behaviour of the raw water tank during normal operation and during the entire attack window. Note that the entire attack window doesn't imply that during this window the plant was under attack but during  different intervals mentioned in Tables\,\ref{tab:validation1} and \ref{tab:validation2}. The graphs in Figure\,\ref{fig:rawWater} represent the behaviour of the raw water tank level during normal operation and during the attack window. The graphs make it clear that there is a significant disruption in the water level during the attack window, thus demonstrating the impact of attack patterns generated. Similarly, we also assessed the impact of an attack on the water level of the Ulta-Filtration (UF) tank during the attack window and normal operation, as shown in Figure \ref{fig:ufWater}. The behaviour of the UF tank is also disrupted due to attacks performed on the plant.

\newcolumntype{C}[1]{>{\centering\let\newline\\\arraybackslash\hspace{0pt}}m{#1}}

\section{Conclusions}
\label{sec:conclusions}
A rule-based machine learning approach is proposed to automatically generate attack patterns for the physical processes in an ICS. The $36$ attack scenarios developed by the domain experts for a water treatment plant were provided as seed to generate the attack patterns. The proposed method generated over 110,000 attack patterns. A three-way validation strategy using invariants, offline datasets, and a live operational testbed was adopted to conduct a thorough validation of the generated attack rules. Based on the results from the case study, we conclude that the ARM-based attack rule generation technique proposed in this work is effective in creating attack patterns and its validation demonstrates only a small proportion of attacks that did not move the plant to an anomalous state.  In the future, we plan to use these attack patterns to conduct the security assessment of ICS and expand the dictionary for signature-based intrusion detection systems. 

\section*{Acknowledgements}
This research is supported by the National Research Foundation, Singapore, under its National Satellite of Excellence Programme “Design Science and Technology for Secure Critical Infrastructure: Phase II” (Award No: NRF-NCR25-NSOE05-0001). Any opinions, findings and conclusions or recommendations expressed in this material are those of the author(s) and do not reflect the views of National Research Foundation, Singapore.

\bibliographystyle{unsrt}
\balance
\bibliography{references.bib}

\end{document}